\documentclass[journal]{IEEEtran}
\IEEEoverridecommandlockouts
\usepackage{cite}
\usepackage{amsmath,amssymb,amsfonts}
\usepackage{algorithmic}
\usepackage{graphicx}
\usepackage{textcomp}
\usepackage{xcolor}
\usepackage{array}
\usepackage{booktabs}  
\usepackage{multirow}  

\usepackage{mathtools}
\usepackage{etoolbox}
\makeatletter
\patchcmd{\@IEEEeqnarray}{\relax}{\relax\intertext@}{}{}
\makeatother

\def\BibTeX{{\rm B\kern-.05em{\sc i\kern-.025em b}\kern-.08em
    T\kern-.1667em\lower.7ex\hbox{E}\kern-.125emX}}
\begin{document}

\title{





Lower Dimensional Spherical Representation of Medium Voltage Load Profiles for Visualization, Outlier Detection, and Generative Modelling

}


\author{
        \IEEEauthorblockN{
                          Edgar Mauricio Salazar Duque,~\IEEEmembership{Student Member,~IEEE,}
                          Bart van der Holst,~\IEEEmembership{Student Member,~IEEE,} \\
                          Pedro P. Vergara,~\IEEEmembership{Member,~IEEE,} 
                          Juan S. Giraldo,~\IEEEmembership{Member,~IEEE,}
                          Phuong H. Nguyen,~\IEEEmembership{Member,~IEEE,}\\ 
                          Anne Van der Molen,~\IEEEmembership{Member,~IEEE,}
                          and~Han (J.G.) Slootweg~\IEEEmembership{Senior Member,~IEEE.}\\
        }

}

\maketitle

\begin{abstract}
This paper presents the spherical lower dimensional representation for daily medium voltage load profiles, based on principal component analysis. The objective is to unify and simplify the tasks for (i) clustering visualisation, (ii) outlier detection and (iii) generative profile modelling under one concept. The lower dimensional projection of standardised load profiles unveils a latent distribution in a three-dimensional sphere. This spherical structure allows us to detect outliers by fitting probability distribution models in the spherical coordinate system, identifying measurements that deviate from the spherical shape. The same latent distribution exhibits an arc shape, suggesting an underlying order among load profiles. We develop a principal curve technique to uncover this order based on similarity, offering new advantages over conventional clustering techniques. This finding reveals that energy consumption in a wide region can be seen as a continuously changing process. Furthermore, we combined the principal curve with a von Mises-Fisher distribution to create a model capable of generating profiles with continuous mixtures between clusters. The presence of the spherical distribution is validated with data from four municipalities in the Netherlands. The uncovered spherical structure implies the possibility of employing new mathematical tools from directional statistics and differential geometry for load profile modelling.
\end{abstract}


\section{Introduction}
The widespread installation of advanced metering infrastructure (AMI) in the medium voltage (MV) distribution grid has enabled distribution network operators (DNOs) to gather enormous volumes of data; within the variety of data collected by the AMI, there are daily load profiles that opens new possibilities for load analysis, forecasting, and load management \cite{wang_review_2019}. These load profiles can be collected at different time resolutions. The resolution refers to the discretisation steps of the time axis at equal intervals, e.g., minutes, quarterly, and hourly. When each discrete step is considered an independent variable, any load profile modelling becomes a multivariate analysis problem. The dimensionality of the model depends on the discretisation resolution. e.g., a daily profile with a quarterly resolution is a 96-dimensional model.

The large number of load profiles collected by the DNOs creates challenges for the multivariate analysis, storing, and efficient use of computing processing power. One technique to aid these challenges is to reduce the dimensionality of the data. In this paper, we refer to dimensionality reduction (DR) as the broad category of linear and nonlinear space embedding methods \cite{guyon_feature_2006}, which reduces the data set to a lower dimensionality space while maintaining most of its information.

In general, four major tasks can benefit from the application of dimensionality reduction (DR) techniques : (i) Clustering can be potentially improved when most of the information is contained in few dimensions \cite{bouveyron_model-based_2014}, (ii) Visualisation and interpretation of the clustering can be made if data is represented in less than three dimensions, (iii) Anomalous readings can be spotted when the data does not follow the structure in the lower-dimensional representation (outlier detection), and (iv) simple generative models can be designed to create load profiles with the same statistical properties as the original data set. For each of the above-mentioned tasks, the data analyst should reach out to different kinds of methods. 

Each task has its own family of methods, which is a considerably large area of research on its own. Here, we mention the relevant literature on the applications of DR for each task, highlighting the relevancy of load profile modelling for power systems.

Clustering techniques are used to understand and allocate consumer behaviour in groups \cite{chicco_load_2021}, which can be used to customise demand response energy efficiency programmes \cite{lin_clustering_2019}. Extensive research has been dedicated to studying the clustering of residential load demand profiles \cite{zhou_review_2013}. Examples of clustering algorithms used are k-means, self-organising maps, hierarchical and spectral clustering \cite{jain_validating_2021}, load decomposition \cite{kwac_household_2014}, c-vine copulas \cite{sun_c-vine_2017}, and probability distribution mixture models \cite{granell_clustering_2015}. It is common to apply these methodologies considering all dimensions and achieving successful results. However, it creates a problem of scalability when the number of profiles increases. Applying the DR technique before clustering is a different approach to improve results and reduce computational time. Principal component analysis (PCA) is generally used as a data preprocessing step as a linear DR technique that preserves most of the variance of the data in a lower-dimensional projection. An important observation from the works that apply DR before clustering is that approximately 85-90\% of the variance for the load profiles can be retained in less than four dimensions using PCA, i.e., \cite{jain_validating_2021, aleshinloye_evaluation_2021, bosisio_method_2020, aleshinloye_performance_2017, chicco_comparisons_2006}. This reduction implies that PCA shows great potential for visualising and analysing clustering results for load profile data with minimal computational effort. However, the study of data structure and lower-dimensional embeddings created by PCA is often neglected. This includes examining common shapes and patterns in the projections of various load profile datasets, which can be valuable for anomaly detection and generative modelling tasks. Combining DR with clustering techniques offers the additional benefit of analysing and visualising the resulting groupings if the reduced dimensions are three or fewer.

Recent research shows applications of nonlinear DR techniques on energy data for visualisation and clustering using neighbour-based techniques, such as lower linear embedding (LLE), Isomap, and t-SNE \cite{arechiga_comparison_2016}. The main drawback of these techniques is the number of parameters needed to tune, e.g., the number of neighbours, components, learning rate, and regularisation, which can significantly affect the projection results. Additionally, nonlinear DR techniques based on neural networks, e.g., UMAP \cite{khan_bottom-up_2023}, and convolutional autoencoders \cite{ryu_convolutional_2020}, show good performance in compression and clustering measure statistics. Nevertheless, creating the non-linear model requires significant time and computational power, often needing data augmentation techniques. Also, they are very flexible models susceptible to random initialisation in the neural network. Consequently, when an autoencoder is trained with a different random initialisation, we can expect different projections or embedding for the same input data. This creates a potential problem in defining anomaly scores in the reduced space and requires reinterpreting the clustering results between tests, restricting its use only to visualisation purposes. In general, nonlinear techniques have great compressibility power. However, they sacrifice stable and consistent projections compared to PCA.

Additionally, DR techniques can also be applied for anomaly detection. Load profile data can have outliers, which are corrupted, abnormal, noisy, and missing data due to various causes. The typical approach to find outliers of the load profile is to use all dimensions and apply kernel smoothing techniques \cite{chen_automated_2010, guo_detecting_2012}, a nearest-neighbours approach using local outlier factor \cite{peng_electricity_2021}, or tree-based methods such as isolation forest \cite{khaledian_real-time_2021}. Those methods are successfully tested and applied for single-meter readings. Usually, outlier detection on load profiles is a laborious process for any data analyst, since an outlier model for each meter should be created. However, in a lower-dimensional space, it is possible can identify outlier profiles from a group of sensors analysing the latent distribution of the data, i.e., probability distribution in the lower-dimensional projection. Simpler models with consistent projections are desired, allowing data analysts to spot anomalous readings quickly.

Generative models can also be developed in a lower-dimensional space to recreate the load profiles from the sensor data. The generative models act as surrogate multivariate probability distribution functions that can be sampled to build large consumption profile databases. Such databases can be used to improve rare event risk assessments \cite{konstantelos_using_2019}, train data-intensive control algorithms such as reinforcement learning \cite{salazar_duque_community_2022}, and evaluate the efficiency of photovoltaic control mechanisms \cite{vergara_comprehensive_2020}. Besides extending a database to arbitrarily large sizes, creating a model in a lower-dimensional space can also compress large volumes of sensor data in a model with few parameters. The more straightforward lower-dimensional data structure from DR techniques leads to simpler generative models for load profile data. 

The four tasks of clustering, visualisation, outlier detection, and generative modelling are highly interlinked in the DR domain. Generally, data analysts address each task separately, requiring an extensive collection of methods as building blocks to create a comprehensive model for load profile analysis. This work proposes projecting load profiles into a simplified lower-dimensional structure\textemdash a sphere. The spherical representation of the data is based on the computationally efficient PCA method, which has the property of consistent projections between different load profile datasets. This property is studied and confirmed with data from four municipalities in the Netherlands. With the spherical representation, we can define simple outlier detection models in the spherical coordinate system. Moreover, the results obtained from unsupervised clustering algorithms, designed to characterise different consumption patterns such as residential, commercial, and mixed zones, can be readily visualised using this spherical representation. We can also create simple generative models based on probability distributions from directional statistics. This spherical model allows us to unify the four previously mentioned topics into one model, offering new research directions under one umbrella.

\subsection{Contributions}
The contributions of this paper are as follows:
\begin{itemize}
    \item We show that standardised daily MV load profiles lie in a hypersphere. Moreover, we showed that applying a PCA-based dimensionality reduction to such standardised profiles results in a sphere projection.
    \item We exploit this spherical model to unify the tasks of clustering, visualisation, anomaly detection, and generative modelling into one concept. This representation allows the development of outlier detection models in the spherical coordinate system, e.g., anomalous patterns and sensors with faulty readings.
    \item We introduce the concept of \textit{order of load profiles} using principal curves, which can be used to perform clustering using a one-dimensional parameter. Based on this ordering concept, a generative model is developed for MV load profiles exploiting the spherical properties of the model, allowing us to control the sampling between clusters in a continuous form.
\end{itemize}

Additionally, the goal of this paper is not to present a model that outperforms all the techniques for each task at the same time, i.e., clustering, anomaly detection, visualisation, and generative modelling, but rather to offer an alternative approach to load profile modelling that can be simpler and practical when dealing with all the tasks at the same time.



\section{Lower dimensional space modelling}
This section aims to create a lower-dimensional spherical representation of the dataset to identify outliers and discuss the mathematical properties of the DR technique. 

The daily load profiles in a large area can be collected in the form of a matrix $\boldsymbol{P} \in \mathrm{R}^{M \times D}$ described as
\begin{IEEEeqnarray}{rCl}\label{eq:profiles_set}
    \boldsymbol{P} &=& \left[\boldsymbol{p}_{1}, \ldots, \boldsymbol{p}_{M} \right],
\end{IEEEeqnarray}
where each row-vector $\boldsymbol{p}$ are the readings for each transformer ($M$), and columns represent the power value of discretised time step ($D$), e.g., for 15min resolution daily profile $D=96$. The idea of DR is to represent such a data set in a decreased $N$-number of variables, i.e. ${N \ll D}$, 


\subsection{Profile Standardisation and Dimensionality Reduction}\label{sec:PCA}
Standardisation of electricity profiles is a common practise for applications concerned with the load profile’s shape rather than the absolute values in consumption. The profile shape depicts the type of consumption in the serviced areas by the MV distribution transformer.

Standardisation of daily MV load profiles is common for applications concerned with the shape of the load profile rather than the absolute values in consumption. The shape depicts the type of consumption in the MV transformer's serviced areas. Normalisation scales the profiles with different active power values into a common range to group profiles with similar shapes.

An essential observation is that standardisation of load profiles brings spherical properties to the dataset in $\boldsymbol{P}$. When we standardise each profile of $\boldsymbol{P}$ using
\begin{IEEEeqnarray}{rCl}
\hat{p}_{i,j}  &=&\frac{p_i - \bar{\mu}_i}{\sigma_i} \quad \forall \quad i=\{1,\ldots,M\},\; j=\{1,\ldots,D\}\label{eq:standardization}\\
\shortintertext{where}
\bar{\mu}_i&=& \frac{\sum_{j=1}^{D} p_{i,j}}{D} \quad \textrm{and} \enspace \sigma_i=\sqrt{\frac{\sum_{j=1}^{D}(p_{i,j}-\bar{\mu}_i)^2}{D}}, \label{eq:mean_std}
\end{IEEEeqnarray}
a standardised matrix $\boldsymbol{\hat{P}}$ is created. It is critical to notice that the distance of the points of $\boldsymbol{\hat{P}}$ with respect to the origin is also constant because
\begin{IEEEeqnarray}{rCl}
||\hat{\boldsymbol{p}}_i||^2  &=& \hat{\boldsymbol{p}}_i \hat{\boldsymbol{p}}_i^{\intercal} \label{eq:euclidean}  \\
                            &=& \hat{p}_{i,1}^2 + \hat{p}_{i,2}^2 + \ldots +\hat{p}_{i,D}^2 \\
\shortintertext{using (\ref{eq:standardization}) and (\ref{eq:mean_std}),}
||\hat{\boldsymbol{p}}_i||^2 &=& \big[  \frac{\sum_{j=1}^D (p_{i,j} - \bar{\mu}_i )}{\sum_{j=1}^D (p_{i,j} - \bar{\mu}_i )/D} \big] = D,\\
||\hat{\boldsymbol{p}}_i|| &=& \sqrt{D}.      \label{eq:norm}               
\end{IEEEeqnarray}

This means that when the dataset profiles $\boldsymbol{P}$ are standardised by rows, the rows have a constant sum and a constant sum of squares. The constant sum of squares means that the sample points in $\boldsymbol{\hat{P}}$ lie on a hypersphere. For simplicity and to keep $||\hat{\boldsymbol{p}}_i||=1$, we normalize the standardised data set as 
%
\begin{IEEEeqnarray}{rCl}
    \boldsymbol{X} &=& \frac{1}{\sqrt{D}} \boldsymbol{\hat{P}}. \label{eq:normalization}
\end{IEEEeqnarray}

\begin{figure}[t]
    \centering
    \includegraphics[width=\linewidth]{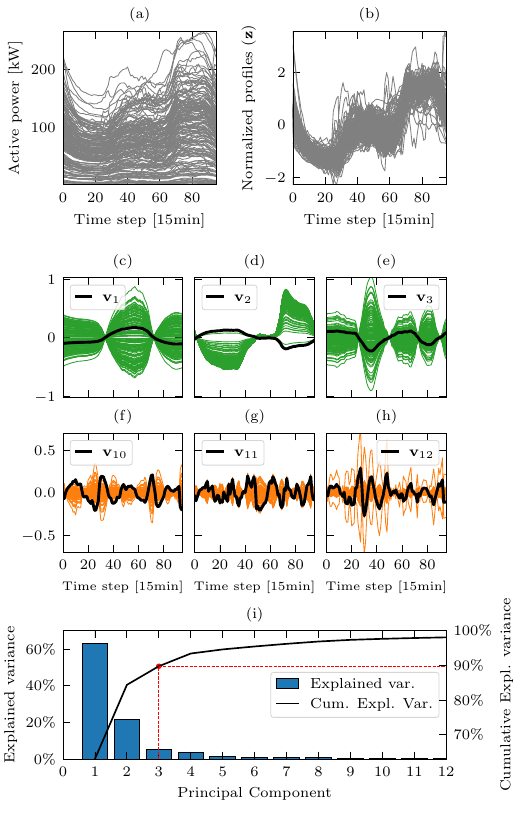}
    \vspace{-10mm}
    \caption{Decomposition of subset of $\boldsymbol{P}$ in its elementary matrices (\ref{eq:elementary}). (a) Original subset $\boldsymbol{P}$. (b) Standardised profiles $\boldsymbol{\hat{P}}$ using (\ref{eq:standardization}). (c)-(e) The first three most significant elementary matrix profiles are green. i.e., $\boldsymbol{X}_1, \boldsymbol{X}_2, \boldsymbol{X}_3$, with their respective eigenvector components in a solid black line. Less significant elementary matrices, i.e., $\boldsymbol{X}_{10}, \boldsymbol{X}_{11}, \boldsymbol{X}_{12}$, are shown in orange, for the eigenvectors (f) $\boldsymbol{v}_{10}$, (g) $\boldsymbol{v}_{11}$, and (h) $\boldsymbol{v}_{12}$. (i) Blue bars show the explained variance by the most important eigenvalues; the solid line is the CEV. The plot is truncated to 12 eigenvalues out of 96. }
    \label{fig:decomposition}
    \vspace{-4mm}
\end{figure}

From now on, we assume that the data set is centered \emph{column-wise}, meaning that the mean of each time step across all load profiles is subtracted from the corresponding column values, i.e., $\boldsymbol{X} \leftarrow \boldsymbol{\textrm{C}}_{M}\boldsymbol{X}$, where $\boldsymbol{\textrm{C}}_{M}$ is the centering matrix for the $M$ samples. The centering process effectively translates the hypersphere so that its new center aligns with the mean of the columns, without changing the overall geometric structure.

Principal component analysis is a linear transformation technique that transforms the data into a new coordinate system so that the variance of the projected data is maximised. This is achieved when the linear transformation comprises the eigenvectors from the covariance matrix of $\boldsymbol{X}$ \cite{bishop_pattern_2006}. Such eigenvectors can be computed using the spectral decomposition of the covariance matrix $\boldsymbol{C} \in \mathbb{R}^{D \times D}$ as
\begin{IEEEeqnarray}{rCl}
    \boldsymbol{C} &=&\frac{1}{M-1} \boldsymbol{X}^\intercal \boldsymbol{X} = \frac{1}{M-1} \boldsymbol{V} \boldsymbol{\Lambda} \boldsymbol{V}^\intercal, \label{eq:covariance}
\end{IEEEeqnarray}
where $\boldsymbol{V}$ is the matrix with the orthonormal eigenvectors (column vectors $\boldsymbol{v}_i \in \mathrm{R}^D$) and $\boldsymbol{\Lambda}$ is a diagonal matrix with the eigenvalues of $S$, i.e., $\boldsymbol{\Lambda}=\text{diag}(\lambda_1, \ldots, \lambda_D)$. The eigenvalues and eigenvectors pairs are sorted so that $\lambda_1 > \lambda_2 > \ldots > \lambda_D$. This sorting specifies each eigenvector's importance order, creating the subspace aligned with the largest data variance. The projection to the new coordinates created by the eigenvectors is given by
\begin{IEEEeqnarray}{rCl}
    \boldsymbol{Z} &=& \boldsymbol{X} \boldsymbol{V}. \label{eq:PCA}
\end{IEEEeqnarray}

The new projected values, $\boldsymbol{Z}$, can be seen as the weights for each eigenvector necessary to recreate $\boldsymbol{X}$. A convenient way to make this clearer for load profile analysis is writing $\boldsymbol{X}$ using the inverse of the transformation of (\ref{eq:PCA}) in terms of the weighted sum of eigenvectors as 
\begin{IEEEeqnarray}{rCl}
    \boldsymbol{X} &=& 
    \boldsymbol{Z}\boldsymbol{V}^\intercal=\begin{bmatrix}
                                                            |   &        |         &        &        |\\ 
                                               \boldsymbol{z}_1 & \boldsymbol{z}_2 & \cdots & \boldsymbol{z}_D\\ 
                                                            |   &        |         &        &        |
                                            \end{bmatrix}
                                            \begin{bmatrix} 
                                                - \, \boldsymbol{v}_1 \, -\\ 
                                                - \, \boldsymbol{v}_2 \, -\\ 
                                                   \vdots    \\ 
                                                - \, \boldsymbol{v}_D  \, -\\ 
                                            \end{bmatrix} \\
      &=& \boldsymbol{z}_1 \boldsymbol{v}_1^\intercal + \boldsymbol{z}_2 \boldsymbol{v}_2^\intercal +
          \cdots + \boldsymbol{z}_D \boldsymbol{v}_D^\intercal\\
      &=& \boldsymbol{X}_1 +\boldsymbol{X}_2 + \ldots + \boldsymbol{X}_D \label{eq:elementary}
\end{IEEEeqnarray}
where matrices $\boldsymbol{X}_i \in \mathbb{R}^{M \times D} \; \forall \: i=\{1,\ldots,D\}$ are elementary matrices. The elementary matrices can be interpreted as the deconstruction of the profiles $\boldsymbol{X}$ in basic profiles ${\boldsymbol{x}_{i,(j,*)} \in \mathbb{R}^D \enspace \forall \: j=\{1,\ldots, M\}}$, which are the row-vector profiles of each elementary matrices. For instance, the subset of residential profiles of dataset $\boldsymbol{P}$ shown in Fig.~\ref{fig:decomposition}(a), has the standardised profiles $\boldsymbol{\hat{P}}$ which shows a clearer residential pattern in Fig.\ref{fig:decomposition}(b). After applying PCA on $\boldsymbol{\hat{P}}$, the three most important eigenvectors (higher eigenvalues) have the elementary matrix profiles plotted in green, and their respective eigenvectors in a solid black line shown in subplots Fig~\ref{fig:decomposition}(c)-(e). Three less significant eigenvectors are shown in Fig.~\ref{fig:decomposition}(f)-(h). We observe that the most important eigenvectors have a lower frequency, carrying the largest variance and capturing the main shape structure of the profiles. On the other hand, the least important eigenvectors capture the higher frequency component of the load profile, resembling noise-like behaviour. We also see that in the case of load profiles, such eigenvectors can be seen as \textit{eigenprofiles}, which are the basic signals needed to rebuild the dataset again.

To determine the number of reduced variables, i.e., the number of \textit{eigenprofiles} to reduce the dataset's dimensionality, we look at the cumulative explained variance ratio (CEV), which is expressed as 
\begin{IEEEeqnarray}{rCl}
    \textrm{CEV}(\lambda_n) &=& \frac{\sum_{j=1}^{n} \lambda_j}{\sum_{k=1}^{D} \lambda_k}.
\end{IEEEeqnarray}

Selecting the $N$ number of major eigenvalues ($\lambda$) means we re-construct the data with the $N$ principal elementary matrices. The bar plot in Fig.~\ref{fig:decomposition}(i) shows the proportion of total variance explained by each eigenvalue, and the solid line is the CEV. It is then shown that 90\% of the data can be explained by the first three \textit{eigenprofiles}. From here, the reduced form for the eigenvector or eigenvalue matrices will have a \textit{hat} notation, e.g., matrix  $\boldsymbol{V}$ that uses only the three principal \textit{eigenprofiles} is referred to as ${\boldsymbol{\bar{V}} \in \mathbb{R}^{D \times 3}}$. 

From the analysis of (\ref{eq:euclidean})-(\ref{eq:norm}), we know that the data set $\boldsymbol{X}$ lies in a hypersphere. Now, if we assume that the data points are distributed along different orthants in the hypershpere, the linear projection of the PCA in \eqref{eq:PCA} using three principal eigenvectors ($\boldsymbol{\bar{V}}$) results at least in an ellipsoid-shaped projection. For simplicity, we will now assume that the projected data in the three-dimensional space have a spherical-shaped structure instead of an ellipsoid. This assumption allows us to apply mathematical tools from circular statistics to perform outlier analysis and generative modelling later in Sections~\ref{sec:order-generative}.

\begin{figure}[t]
    \centering
    \includegraphics[width=\linewidth]{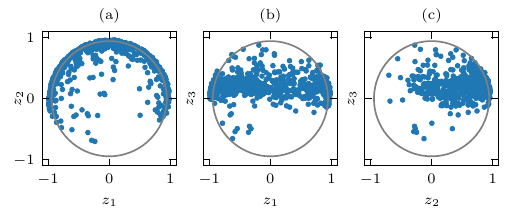}
    \vspace{-7mm}
    \caption{The values of the projection $\boldsymbol{Z}$ in a 3-dimensional space. (a-c) Orthographic projection of the sphere. Each blue point represents a single transformer's daily profile. The sphere overlayed in the data is found via (\ref{eq:optimization}).}
    \label{fig:sphere}
    \vspace{-4mm}
\end{figure}

To visually confirm the spherical-shaped structure of the projected data, Fig.~\ref{fig:sphere}  shows the results of the PCA using three principal components for daily profiles from 560 MV distribution transformers from a municipality in the Netherlands. Each blue point in three dimensions corresponds to a \emph{lower-dimensional} representation of the daily load profile of 96 dimensions (15-minute resolution daily profile).

Interestingly, most projected data points are agglomerated on one side of the sphere revealing a lower dimensional latent distribution, which forms an arc-shaped structure, clearly seen in Fig.~\ref{fig:sphere}(a). Some points lie outside of the latent distribution, meaning they represent a weighted sum of \textit{eigenprofiles} that are not usual from the dataset, creating an outlier-type of profiles. The interest is in labelling those outliers to study the source for such anomalous behaviour.

\section{Sphere Modelling and Outlier Detection}\label{sec:outlier}
After the previous findings, the natural step is to create a sphere as a mathematical model that helps us visualise and discern whether a profile is an outlier. The sphere parameters are found by solving the following optimisation problem 
\begin{IEEEeqnarray}{c}
    \min_{\boldsymbol{c}, \rho} \quad \sum_{j=1}^{3}\sum_{i=1}^{M} (z_{i,j} - c_j)^2 - \rho, \label{eq:optimization}
\end{IEEEeqnarray}
using the change of variable $\rho=r^2$ for the radius, and vector $\boldsymbol{c}= [ c_1, \ldots, c_r]$ as the sphere's centroid. The solution of the sphere parameters allows us to centre the dataset around the origin. i.e. $\boldsymbol{z} \leftarrow \boldsymbol{z} - \boldsymbol{c}$, to apply a change of coordinate system to build models for outlier detection. We convert centred points to the spherical coordinate system using
\begin{IEEEeqnarray}{rCl}
    \varphi_i &=& \textrm{sgn}(\hat{z}_{2,i}) \arccos\big(\hat{z}_{1,i}/ \sqrt{\hat{z}_{1,i}^2 + \hat{z}_{2,i}^2 }\big),\\
    \theta_i &=&  \arccos(\hat{z}_{3,i}/ \sqrt{\hat{z}_{1,i}^2 + \hat{z}_{2,i}^2 + \hat{z}_{3,i}^2}),\\
    r_i &=& \sqrt{\hat{z}_{1,i}^2 + \hat{z}_{2,i}^2 + \hat{z}_{3,i}^2}, \qquad \forall \; i=\{1,\ldots,N\}.
\end{IEEEeqnarray}

\begin{figure}[t]
    \centering
    \includegraphics[width=\linewidth]{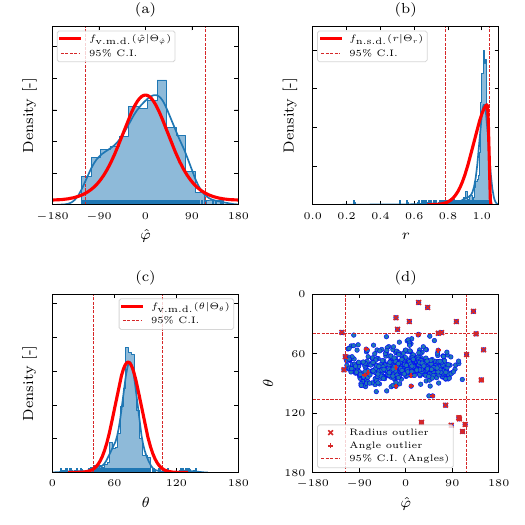}
    \vspace{-10mm}
    \caption{Probability distributions of the spherical projection variables for the dataset $\boldsymbol{X}$. (a) Azimuthal angle distribution centred around the mean. (b) Radius. (c) Polar angle. (d) 2D projection using angle values. Flagged points in red are outliers, which are the points that fall outside the rejection region created by the 95\% confidence interval (CI) from each fitted distribution (delineated by the vertical and horizontal red dotted lines).}
    \label{fig:anomaly_dist}
    \vspace{-4mm}
\end{figure}

The azimuthal angle $\varphi_i$ has the positive $x$-axis as reference ($z_1$), and the polar angle $\theta_i$ is measured from the positive $z$-axis ($z_3$). The coordinate system change makes it easier to fit parametric distributions for each variable as shown in Fig.~\ref{fig:anomaly_dist}, where the angle $\varphi$ is centered around its mean, $\hat{\varphi}=\varphi-\bar{\varphi}$, to aid the visualisation. For the angles, we fit a von Mises distribution (VMD) as a continuous probability distribution used to describe angles in circular statistics as shown in Fig.~\ref{fig:anomaly_dist}~(a), and (c). The radius variable is modelled with a skew-normal distribution (SND) as the data is concentrated on a thin shell of the sphere, creating a skewed distribution (Fig.~\ref{fig:anomaly_dist}~(b)). The parameters for each distribution, i.e., $\Theta_{\hat{\varphi}}$, $\Theta_{\theta}$,  $\Theta_{r}$, are found via maximum likelihood estimation. 

The outlier profiles are deduced using the rejection region created from the 95\% confidence interval for each fitted distribution shown as vertical dotted red lines in Fig.~{\ref{fig:anomaly_dist}}. The flagged data points are discriminated between outliers from the angle models ($\varphi_i$ and $\theta_i$) and radius ($r$) as shown in Fig.~\ref{fig:anomaly_dist}(d). Details about the characteristics of those outliers are discussed in the case study in Section~\ref{sec:case-study}.

\section{Ordering and Generative Modelling}\label{sec:order-generative}
This section dives into the load profiles' lower dimensional structure. It discusses the arc-shape pattern in the sphere shown in Fig.~\ref{fig:sphere}(a), suggesting that the load profiles of a large area have a latent ordering, which can be used for clustering and generative modelling purposes.

\subsection{Ordering by Principal Coordinate Analysis}\label{sec:ordering}
The proximity of the points in the sphere on Fig.~\ref{fig:sphere} implies a great degree of similarity between the load profiles. Meaning that the profiles are constructed with similar weights ($z$-values) combinations of \textit{eigenprofiles}. Such similarity between load profiles can be quantified using the dot product between them, creating a similarity matrix of the form
\begin{IEEEeqnarray}{rCl}\label{eq:similarity_matrix}
    \boldsymbol{S} = \boldsymbol{X} \boldsymbol{X}^\intercal = \begin{bmatrix}
                         \cos{(\gamma_{1,1})}      & \ldots     &      \cos{(\gamma_{1,M})}\\ 
                         \vdots                    & \ddots     &      \vdots               \\ 
                         \cos{(\gamma_{M,1})}      & \ldots     &      \cos{(\gamma_{M,M})}       
                        \end{bmatrix},
\end{IEEEeqnarray}
this matrix is also called the Gram matrix. The dot product between vectors (profiles) of $\boldsymbol{X}$ is determined by $\boldsymbol{x}_i \boldsymbol{x}_j^{\intercal} = ||\boldsymbol{x}_i|| \cdot ||\boldsymbol{x}_j|| \cos{(\gamma_{i,j})}$, and knowing from (\ref{eq:normalization}) that the profiles are normalised vectors, i.e., $||\boldsymbol{x}||=1$, the similarity matrix is reduced to the cosines of the angles between the vectors, ranging between $[-1,1]$. The dimensionality of the similarity matrix $S \in \mathbb{R}^{M \times M}$ is high for large samples ($M \gg D$). Therefore, it is required to apply a DR technique to represent $\boldsymbol{S}$ in a low-dimensional space. Reducing similarity matrices into a lower dimensional projection is common in biology, archaeology, and psychology using multidimensional scaling techniques (MDS). 


Principal Coordinate Analysis (PCoA), or classical MDS, is when the PCA technique is applied to the double-centred Gram matrix, i.e., $\boldsymbol{G} = \boldsymbol{\textrm{C}}_{M} \boldsymbol{S} \boldsymbol{\textrm{C}}_{M}$, which has a spectral decomposition $\boldsymbol{G} =\boldsymbol{\xi} {\Lambda} \boldsymbol{\xi}^{\intercal}$. The objective of PCoA is to preserve the similarity structure (distances between points) in a lower dimensional space. 

When the matrix $\boldsymbol{S}$ has a \emph{band matrix} structure, the PCoA projection has an interesting arc-shaped effect pattern \cite{morton_2017}, like the one shown in Fig.\ref{fig:sphere}(a) for the load profiles. A \emph{band matrix} in our context, means the highest similarity values are closer to the diagonal. The arc effect has been studied and discussed in other study fields in the past for different types of similarity matrices \cite{podani_2002}, and is often seen when the sample set $\boldsymbol{X}$ has an underlying latent ordering \cite{kendall_1970}; or the data comprises a gradual rate of change of a process \cite{campbell_2018}. The parameterisation of these lower dimensional arcs or paths can be used to recover the underlying ordering from the dataset. In our work, knowing the load profile order helps us to determine threshold values to determine clusters for different consumption types and create generative models in the lower dimensional space.

\begin{figure}[t]
    \centering
    \includegraphics[width=\linewidth]{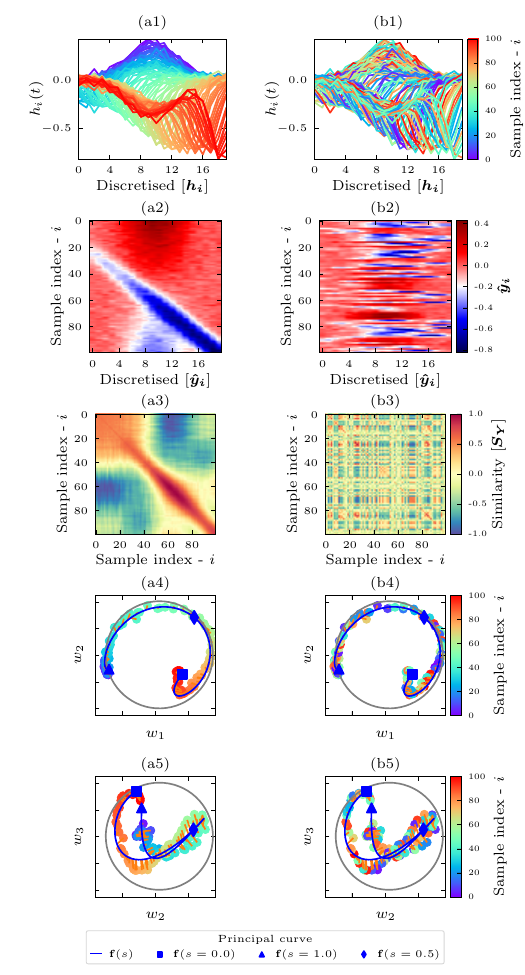}
    \vspace{-8mm}
    \caption{Example of latent space ordering for the process in (\ref{eq:process}). (a1) Data matrix $\boldsymbol{H}$ created by discretisation of (\ref{eq:process}) using 20 time steps. (a2) Heatmap for standardised matrix $\boldsymbol{\hat{Y}}$. (a3) Heatmap similarity matrix $\boldsymbol{S_{\boldsymbol{Y}}}$. (a4) and (a5) spherical orthographic projection for the PCoA applied to $\boldsymbol{S_{\boldsymbol{Y}}}$ using three principal components which shows a clear arc-shape structure. (b1) $\boldsymbol{H}$ with shuffled rows. (b2) The gradual change pattern is lost due to shuffling. (b3) The banded structure of $\boldsymbol{S_{\boldsymbol{Y}}}$ is absent. (b4) and (b5) The latent ordering of the samples still exists, and the original sample labels ($i$) can be recovered following the parametrised curve that passes through the middle of the points.}
    \label{fig:order_example}
    \vspace{-4mm}
\end{figure}

\subsubsection*{Illustrative example for PCoA}
As an illustrative example of how PCoA is used, consider the following non-linear function that gradually changes for each sample:
\begin{IEEEeqnarray}{rCl}
h_i(t)    &=& \tau(i) g(t, 1) + (1 - \tau(i)) g(t, \nu(i)) + 0.03 \; \varepsilon \label{eq:process}\\
g(t, \mu) &=& \frac{1}{\sqrt{2\pi}}  \exp{(-\frac{1}{2}(t-\mu))} \IEEEyessubnumber \\
\nu(i)    &=& 0.08 i - 4 \IEEEyessubnumber \\
\tau(i)   &=& 0.02 i - 1 \qquad \forall \: i=\{1, \ldots, 100\},  \IEEEyessubnumber \\
\varepsilon &\sim& \mathcal{N}(0,1) \IEEEyessubnumber          
\end{IEEEeqnarray}
variable $i$ is the sample index, and each sample profile $h_{i}(t)$ is discretized by 20 steps, creating a row-vector $\boldsymbol{h}_i \in \mathbb{R}^{20}$. The row-vectors builds the data matrix $\boldsymbol{H} = [ \boldsymbol{h}_1, \ldots, \boldsymbol{h}_{100}] \in \mathbb{R}^{100 \times 20}$. The samples $\boldsymbol{h}_i$ are labelled and coloured in the increasing sample index order shown in Fig.~\ref{fig:order_example}(a1). The gradual change can be seen as a heatmap in Fig.~\ref{fig:order_example}(a2) for the standardised row-vectors applying (\ref{eq:standardization}), i.e., $\boldsymbol{H} \xrightarrow{\mathrm{std}} \boldsymbol{\hat{Y}}$.

Using (\ref{eq:normalization}), i.e.,  $\boldsymbol{\hat{Y}} \xrightarrow{\mathrm{norm}} \boldsymbol{Y}$ and computing the similarity matrix, $\boldsymbol{S_{\boldsymbol{Y}}}=\boldsymbol{Y} \boldsymbol{Y}^\intercal$, a clear band matrix structure is appreciated in Fig.~\ref{fig:order_example}(a3). The PCoA over the double centered Gram matrix $\boldsymbol{G_{Y}}$, and using the three principal eigenvectors, i.e., $\boldsymbol{W}=\boldsymbol{\bar{\Lambda}}^{\frac{1}{2}} \boldsymbol{\bar{\xi}}$, is shown in Fig.~\ref{fig:order_example}(a4),(a5) with a spherical orthographic projection. The projections show a clear arc-shaped latent ordering, i.e., the colours of the sample index follow the arc in a clockwise motion in Fig.~\ref{fig:order_example}(a4). 

The second column of Fig.~\ref{fig:order_example} shows the same procedure, only that this time the sample index is shuffled, i.e., the row-vector of $\boldsymbol{H}$ are shuffled (Fig.\ref{fig:order_example})(b1). This is more likely the real case scenario for electricity load profiles, where we do not know the correct order beforehand of similarity between areas serviced by MV transformers. Notice that without order, there is no discernible pattern change in Fig.\ref{fig:order_example}(b2), nor a band-matrix structure as shown in Fig.\ref{fig:order_example}(b3). However, the latent arc shape remains unchanged as shown in Fig.\ref{fig:order_example}(b4). To recover the ordering, we create a parametrised curve that passes through the middle of the lower dimensional points. Then, we re-label the samples following the projection of the points to the curve.

The reason for choosing the dot product as the similarity matrix over other types of distance matrix options is because reducing (\ref{eq:similarity_matrix}) and (\ref{eq:normalization}) produce the same lower dimensional projection \cite{cox_2008}. Nevertheless, it is helpful to compute the matrix $\boldsymbol{S}$ after the sample ordering procedure to visualise and verify the existence of a band matrix structure.

In the case of load profiles data, the arc-shape pattern can be seen in Fig.~\ref{fig:sphere}(b), indicating a latent ordering of the dataset. The following subsection discusses the procedure for finding the parametrised curve.

\subsection{Principal Curve Model}
The principal curve is a technique to summarise complex $N$-dimensional distributions into one dimension \cite{hatie_1989}. The technique consists in creating a parametric curve $\boldsymbol{f}(s)$ that passes through the middle of the data points in a smooth way. Here, we use this concept to find the latent ordering for the load profiles. The curve $\boldsymbol{f}(s)$ comprises $N$ functions with a single variable $s$. Precisely is defined as
\begin{IEEEeqnarray}{rCl}\label{eq:principal_curve}
    \boldsymbol{f}(s) &=& [f_1(s), \ldots, f_N(s)]^\intercal, \quad s \in [0,1].
\end{IEEEeqnarray}
Each function $f_k(s) \mapsto \mathbb{R} \; \forall \, k=\{1,\ldots,N\}$ can be parametrised by nearest-neighbours, kernel or spline smoothers. In this study, we used splines due to the smooth spherical surface, and the number of functions is three ($N$=3). i.e., one function per each cartesian coordinate. A general procedure to find $\boldsymbol{f(s)}$ is described in \cite{hatie_1989}. However, a more robust version is developed in \cite{salazar_2023}, where the initial iteration exploits the fact that most data reside in a shell in a sphere, increasing the correct convergence likelihood. The iterative algorithm's objective is to find the spline parameters that minimise the orthogonal projections of the data into the curve. The projection index is defined as
\begin{IEEEeqnarray}{rCl}\label{eq:projection}
    \lambda_{\boldsymbol{f}^{(\cdot)}}(\boldsymbol{z}) &=& \sup_{s}\{ s: ||\boldsymbol{z} - \boldsymbol{f}(s)|| = \mathop{\rm inf}\limits_{\mu} || \boldsymbol{z} - \boldsymbol{f}(\mu)||\}
\end{IEEEeqnarray}
The projection index $\lambda(\cdot)$ of the PCA point $\boldsymbol{z_i}$ is the $s_i$ value for which $\boldsymbol{f}(s_i)$ is closest to $\boldsymbol{z_i}$, i.e., $\lambda(\boldsymbol{z_i}) \mapsto s_i$. This means that each point in the sphere, representing a profile, has a corresponding $s$-value between 0 and 1 for the parametrised curve. For instance, the curve fitted for the points in Fig.\ref{fig:order_example}(a4,b4,a5,b5) shown as a solid blue line, is used to sort the projected values of $\boldsymbol{w}$ on the curve. Once the $\boldsymbol{w}$ values are sorted, we recovered the original ordering of the process in (\ref{eq:process}).

Using a single-valued variable $s$ to represent profiles can aid in creating data clusters by dividing it into groups or bins, with each bin representing a cluster. 

\subsection{Generative Modelling}
The benefit of having the principal curve is combining it with a normal probability distribution to create a generative model. In the field of directional statistics, the generalisation of the normal probability distribution for a $(N-1)$-sphere is the von-Mises-Fisher (VMF) distribution. For the specific case of a three-dimensional sphere, the VMF distribution is described as
\begin{IEEEeqnarray}{rCl}\label{eq:von_misses}
    \boldsymbol{f}_{\mathrm{vmf}}(\boldsymbol{z}; \boldsymbol{\mu}_\mathrm{vmf}, \kappa) 
    &=& \frac{\kappa}{2\pi(e^{\kappa} - e^{-\kappa})}\exp{(\kappa \boldsymbol{\mu}_\mathrm{vmf}^\intercal \boldsymbol{z})},
\end{IEEEeqnarray}
where $\kappa$ is the concentration parameter and controls the spread of the samples over the sphere. Parameter $\boldsymbol{\mu}_{\mathrm{vmf}} \in \mathbb{R}^3$ is the directional vector where the normal distribution is centred. Here, we use $\boldsymbol{f}(s) \equiv \boldsymbol{\mu}_{\mathrm{vmf}}$, to dynamically and continuously move the VMF distribution over the sphere creating data from the desired section. The synthetic standardised shape profiles $\boldsymbol{\bar{P}}$ can be obtained by sampling from (\ref{eq:von_misses}) and using the inverse process described in the previous sections as
\begin{IEEEeqnarray}{rCl}\label{eq:synthetic}
    \boldsymbol{\bar{Z}} &\sim&  \boldsymbol{f}_{\mathrm{vmf}}(\cdot;\kappa, \boldsymbol{f}(s))  \mapsto  \boldsymbol{\bar{X}} = \boldsymbol{\bar{Z}}\boldsymbol{\bar{V}}^\intercal \mapsto  \boldsymbol{\bar{P}}=\sqrt{D}\boldsymbol{\bar{X}}
\end{IEEEeqnarray}
\section{Case study}\label{sec:case-study}
This section discusses (i) The meter readings labelled as outliers in Section~\ref{sec:outlier} and the visualisation of load profile clustering results on the sphere. (ii) The ordering of load profiles created by the principal curve techniques. (iii) Generative modelling benefits and limitations, and (iv) generalisation of the spherical modelling for other MV load profile datasets. Results presented in this Section use the profile dataset introduced earlier (560 MV load profiles) to keep consistency between the results presented in the previous sections.

\begin{figure}[t]
    \centering
    \includegraphics[width=0.7\linewidth]{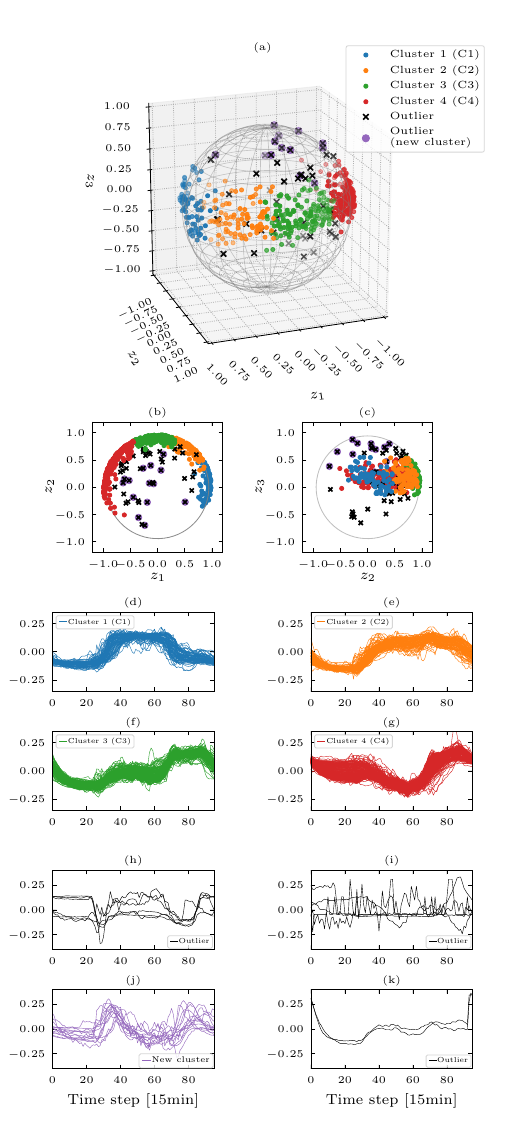}
    \vspace{-6mm}
    \caption{Clustered profiles from Municipality 1 and outlier identification. (a-c) Visualisation for clustering results and outliers using spherical modelling. Daily Profiles that are not considered outliers are grouped into four: (d) Commercial/Offices (e) Mix of residential and commercial (f) Residential (g) Residential with high PV penetration. (h-k) Anomalous reading labelled by outlier models from section \ref{sec:outlier}. (j) The scatter points on the top of the sphere marked as outliers are a cluster of their own.}
    \label{fig:sphere_clusters}
    \vspace{-7mm}
\end{figure}

\subsection{Profile Clustering and Anomalous Meter Readings}
The outlier meter readings labelled by the rejection region from the probability distributions presented in Section~\ref{sec:outlier} are shown with \textit{x} markers in Fig.~{\ref{fig:sphere_clusters}}(a-c). It has been found that from the three spherical variables, the radius is the most effective descriptor of the defective meters, and its labelled profiles are shown in Figure~{\ref{fig:sphere_clusters}}(h,i,k), where subplot (i) shows 5 profiles for meters that are recording noise. 

\begin{figure*}[t]
    \centering
    \includegraphics[width=\linewidth]{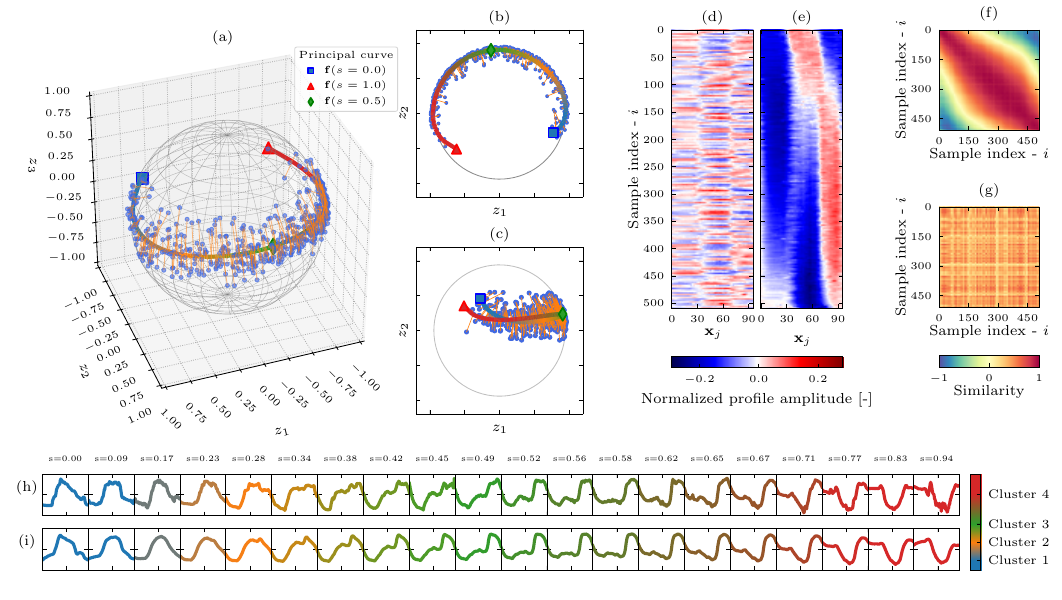}
    \vspace{-10mm}
    \caption{Ordered dataset for Municipality 1. (a-c) The principal curve model passes through the middle of the data points representing the load profiles. It shows a latent ordering due to the compactness of the cloud of points. Orange lines are the orthogonal projection of the points to the principal curve. (d) Heat map of the original load profile data set matrix $\boldsymbol{P}$. (e) The ordered matrix $\boldsymbol{P}$ using the principal curve uncovers the latent ordering of the dataset, confirmed in (f) by the banded similarity matrix. (g) The disorganised original dataset $\boldsymbol{P}$ produces a non-structured similarity matrix. (h) Gradual change between clusters is exposed by plotting twenty original profiles ordered by the principal curve. The change goes from commercial areas (blue) to residential areas with high PV penetration (red). (i) The same twenty profiles are recovered using its orthogonal projection to the principal curve.}
    \label{fig:principal_curve}
    \vspace{-4mm}
\end{figure*}

A more challenging failure to detect is shown in Figure~{\ref{fig:sphere_clusters}}(h), where 6 meters are only storing absolute values of active power, meaning that the power injection from PV panels was recorded as consumption and not generation, explaining the increased consumption bump seen in daylight hours. This failure is challenging because outlier detection techniques that rely on past data, like smoothing kernels \cite{chen_automated_2010, guo_detecting_2012}, can not detect similar and repetitive failure readings. Nevertheless, with the spherical modelling, those profiles stand out from the latent distribution of the complete load profile set, making its identification easier. The profiles shown in Fig.~\ref{fig:sphere_clusters}(k) are two large areas with midnight operations. These profiles are in fact, outliers, as their unique shape are only 2 out of 560, but do not correspond to meter failure. 

Profiles labelled as outliers by the azimuth and polar angle distribution models are grouped in Fig.~\ref{fig:sphere_clusters}(j), which are the points scattered at the top of the sphere (purple circle markers). These profiles have a pronounced high load consumption early in the morning, and PV generation creates a valley in the middle of the profile. This particular shape structure requires higher weight ($\boldsymbol{z_3}$) for the third \textit{eigenprofile} shown in Fig.~{\ref{fig:muni_distributions}(c) (Municipality 1)}, to recreate the morning load spike consumption. This group of 14 profiles can be considered a cluster independently and are not defective meter readings. 

We excluded the data points associated with defective meters on the sphere and employed various clustering techniques, namely k-means, Gaussian mixtures, Hierarchical, and Spectral clustering. It should be noted that none of the clustering techniques could detect the profiles in Fig.~\ref{fig:sphere_clusters}(k) as a unique cluster, as they incorporated those points into different groups. Therefore, we isolated those readings from the clean profiles set to repeat the clustering for a total of 533.

The coloured visualisation for the Agglomerative Hierarchical clustering (AggHC) results with Ward distance is shown in Fig.~{\ref{fig:sphere_clusters}}(a-c). Four representative groups are found and correspond to different consumption activity zones. From Fig.~{\ref{fig:sphere_clusters}} subplot (d) is C1: Commercial/Offices, (e) C2: Mix of residential and commercial, (f) C3: Residential, and (g) C4: Residential with high PV penetration. The spherical visualisation of the clusters offers a different view of the proximity between clusters in the lower dimensional domain. The clear case is for the cluster with a mixture of residential and commercial zones (C2). This cluster lies between commercial and residential areas clusters, indicating that C2 is a transition section on the sphere between the adjacent clusters. Likewise, C3 is viewed as the transition region between C4 and C2.

\subsection{Latent Ordering by Principal Curve}\label{sec:latent_order}
The principal curve fitted in the cleaned dataset after clustering is shown in Fig.~\ref{fig:principal_curve}(a-c). The orange lines represent the orthogonal projections of each data point into the principal curve, i.e., $\lambda(\boldsymbol{z}_i) \mapsto s_i \mapsto \boldsymbol{f}(s_i)$. Figure~\ref{fig:principal_curve}(h) shows 20 profiles from the matrix (\ref{eq:normalization}) ordered by $s_i$ values, which gives a clear view of how the profiles gradually change between clusters, confirming a latent ordering for the profiles set. The projected points in the curve can also be transformed back into standardised values, i.e.,~${\boldsymbol{\bar{P}} = \sqrt{D}\boldsymbol{f}(s_i)\boldsymbol{\bar{V}}^\intercal}$, and is shown in Figure~\ref{fig:principal_curve}(i). These transformed profiles have a smoother shape than the original profiles because the reconstruction is done using the three principal \textit{eigenprofiles}, and the high-frequency components, characterised by the least important \textit{eigenprofiles}, are not used as discussed in Section~\ref{sec:PCA}. 

The heat map in Fig.~\ref{fig:principal_curve}(d) shows the matrix (\ref{eq:profiles_set}) as it was initially collected from the AMI database, without any particular order, and its similarity matrix (\ref{eq:similarity_matrix}) is plotted in Fig.~\ref{fig:principal_curve}(g). Both heat maps do not show any structure. After re-ordering (\ref{eq:profiles_set}) with the principal curve, Fig.~\ref{fig:principal_curve}(e) depicts a gradual change in the profiles. Also, the similarity matrix of the ordered dataset, in Fig.~\ref{fig:principal_curve}(f), shows a banded structure, explaining the arc-shape pattern discussed in Section~{\ref{sec:ordering}}.

The $s_i$ is a single-valued descriptor for each ordered profile. In our case, $s=0.0$ corresponds to the first sample, $\boldsymbol{f}(s=0.0)$ in a blue square marker in Fig.~\ref{fig:principal_curve}(a); and $s=1.0$ for the last sample, $\boldsymbol{f}(s=1.0)$ in a red triangle marker in the same subplot. The principal curve can be split into segments to create hard clusters over the dataset. Each segment can be seen as bins to group the profiles. In this example, the bins range that concur with AggHC results are $\textrm{C1} \in [0.0,0.2)$,  $\textrm{C2} \in [0.2, 0.4)$,  $\textrm{C3} \in [0.4, 0.6)$ and  $\textrm{C4} \in [0.6, 1.0]$. 

The ordering with the principal curve brings two significant advantages compared to other clustering techniques: (i) The profiles within each cluster are ordered, which gives the DNO additional information about ranked areas within the city based on their similarities. Consequently, this data offers detailed insights into the synchronisation of peak and valley consumption times in these areas. (ii) Principal curve conveys a sense of continuity between clusters, giving extra information to quantify mixtures of areas. For instance, the centroid of C1, i.e., $s_{c1}=0.1$, is the representative value for commercial areas, and C3 centroid, i.e., $s_{c3}=0.5$ for the residential areas. The profiles that lie in the curve segment created by $[s_{c1}, s_{c3}]$ can be treated as a mixture between both activity areas; for example,  $s=0.22$ would be a profile for an area with 30\% residential and 70\% commercial consumption.

\subsection{Generative Modelling}
The principal curve acts as the probability distribution centre ($\boldsymbol{\mu}^{\intercal}_{\mathrm{vmf}}$) in the model (\ref{eq:von_misses}), allowing the generation of points in different sections on the sphere depending on the variable $s$. The spherical generative model, labelled as (VMF), is tested by creating (sampling) 10 profiles using (\ref{eq:synthetic}) for 25 points over the principal curve. i.e., $\boldsymbol{\bar{Z}}_k \sim \boldsymbol{f}_{\mathrm{vmf}}(\cdot;\kappa, \boldsymbol{f}(s_k))$, where $s_k = s_{k-1} + \Delta{s}, \, \Delta{s}=1/25, \, s_0=0.0, \; \forall \; k=\{1, \ldots, 25\}$. The concentration parameter $\kappa=7.1$ is taken from the polar angle model parameters ($\Theta_{\theta}$). The total generated profiles are 250, and using their projected $s$-values into the principal curve, they are clustered using the bin values from Section~\ref{sec:latent_order}. 

\begin{figure}[t]
    \centering
    \includegraphics[width=\linewidth]{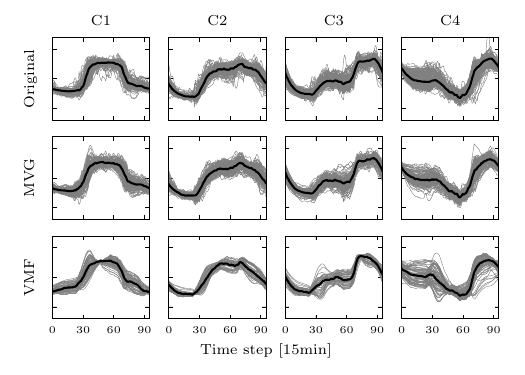}
    \vspace{-10mm}
    \caption{Comparison between original and synthetic profiles generated by spherical modelling (\ref{eq:von_misses}), labelled as VMF and MVG. The solid black line represents the median for the profiles in each cluster.}
    \label{fig:synthetic}
    \vspace{-4mm}
\end{figure}

\begin{table}[t]
\centering
\caption{Generated load profiles comparison metrics for each cluster}\label{table:dist_metrics}
\begin{tabular}{llrrrr}
\toprule
  Metric   &  Model   &      C1 &      C2 &      C3 &     C4 \\
\midrule
\multirow{2}{*}{KS}   & VMF &   .054 &   .032 &   .031 &  .034 \\
                      & MVG &   .033 &   .022 &   .020 &  .023 \\ \midrule
\multirow{2}{*}{RMSE} & VMF &   .022 &   .019 &   .018 &  .020 \\ 
                      & MVG &   .008 &   .005 &   .007 &  .010 \\
\bottomrule
\end{tabular}
\vspace{-4mm}
\end{table}

For comparison purposes, we generated the same number of profiles per cluster using a multivariate Gaussian distribution (MVG), shown in \cite{duque_conditional_2021} as a suitable model to generate MV load profiles. The synthetic profiles from the models are plotted in Fig.~\ref{fig:synthetic} and compared against the original dataset. It is observed that the profiles generated by VMF are smoother than the original profiles due to the limited reconstruction of the profile with three components. This could limit the VMF application for analyses that require high accuracy for active power values every fifteen minutes. This is expected to be less critical for profiles recorded at lower resolutions, e.g., hourly. However, the main VMF model advantage is that it can control the generation of profiles between clusters, as shown in Fig.~\ref{fig:principal_curve}(i). This is not feasible with any other mixture model technique that typically treats each cluster as a categorical variable (discrete). The profile quality is quantified in Table~\ref{table:dist_metrics} by using the Kolgomorov-Smirnov (KS) probability distance metric, and the root mean squared error (RMSE) between the original and the synthetic data, which shows that VMF and MVG can closely model the original profiles.

\subsection{Spherical Modelling for Different Datasets}
Most of the results in this paper are based on data from one municipality in the Netherlands (Municipality 1). Even though the mathematical formulation of the spherical modelling is robust, the natural question arises about generalising the model for different MV load profile datasets and, specifically, inquiring if other datasets show the same latent distribution (concentrated cloud of points in a sphere). Here, we use the profiles from four different municipalities in the Netherlands. Instead of repeating all figures and analyses on each one, we argue that all datasets have similar properties in the latent space, and spherical modelling can be applied to any of them.

\begin{figure}[t]
    \centering
    \includegraphics[width=\linewidth]{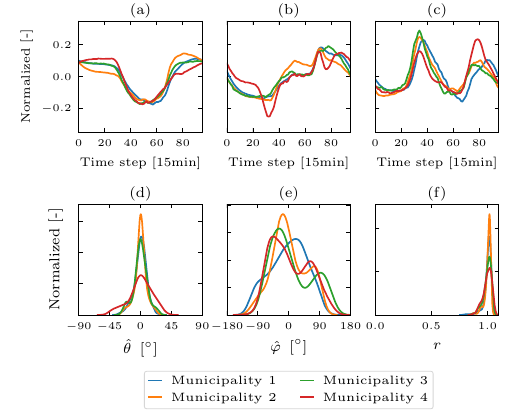}
    \vspace{-8mm}
    \caption{Data properties for four different municipalities in the Netherlands. (a,b,c) Three most important \textit{eigenprofiles}. Probability distributions for the spherical coordinates variables: (d) Polar and (e) Azimuth angles (centred around the mean), and (f) radius. The datasets show similar \textit{eigenprofiles} and the lower dimensional spherical coordinates characteristics. }
    \label{fig:muni_distributions}
    \vspace{-4mm}
\end{figure}

\begin{table}[t]
\centering
\caption{Cumulative Explained Variance (\%) of Principal Components for Different Municipalities in the Netherlands}\label{table:cev_municipalities}
 \scalebox{0.93}{
\begin{tabular}{lrrrrrrr}
\toprule
Municipality &     PC1 &      PC2 &      PC3 &      PC4 &     PC5 
    & \begin{tabular}[c]{@{}r@{}}Transf. \\ count\end{tabular} 
    &  \begin{tabular}[c]{@{}r@{}}Outilers \\ (\%)\end{tabular}  \\
\midrule
Municipality 1 &  68.7 &  88.4 &  92.5 &  95.1 &  96.0 & 560 & 27 (5.0\%) \\
Municipality 2  &  69.9 &  88.3 &  92.1 &  94.3 &  95.9 & 375 &  9 (2.4\%)\\
Municipality 3   &  71.8 &  87.1 &  91.0 &  93.4 &  94.5 & 354 & 29 (8.2\%)\\
Municipality 4  &  61.1 &  73.0 &  81.8 &  85.8 &  88.2 & 392 & 61 (16.9\%)\\
\bottomrule
\end{tabular}}
\vspace{-4mm}
\end{table}

\begin{table}[t]
\centering
\caption{Spherical Modelling Summary Statistics}\label{table:statistics_sphere}
\begin{tabular}{clrrrr}
\toprule
  Variable     &  Municipality   &  Mean &     Std &  Skewness &  Kurtosis \\
\midrule
\multirow{4}{*}{$\hat{\theta}$} & Municipality 1 &       0.000 &   9.149 &    -0.512 &     0.960 \\
       & Municipality 2 &       0.000 &   8.660 &    -0.321 &     1.396 \\
       & Municipality 3 &       0.000 &  10.039 &    -0.181 &     0.615 \\
       & Municipality 4 &       0.000 &  17.236 &    -0.449 &    -0.013 \\ \cmidrule(lr){1-6}
\multirow{4}{*}{$\hat{\varphi}$} & Municipality 1 &       0.000 &  52.811 &    -0.229 &    -0.787 \\
       & Municipality 2 &       0.000 &  49.195 &     0.146 &    -0.674 \\
       & Municipality 3 &       0.000 &  58.365 &     0.543 &    -0.989 \\
       & Municipality 4 &       0.000 &  55.126 &     0.347 &    -1.092 \\ \cmidrule(lr){1-6}
\multirow{4}{*}{$r$} & Municipality 1 &       1.002 &   0.036 &    -2.640 &     9.205 \\
       & Municipality 2 &       1.003 &   0.029 &    -2.221 &     5.773 \\
       & Municipality 3 &       0.999 &   0.036 &    -1.203 &     1.384 \\
       & Municipality 4 &       0.998 &   0.039 &    -0.815 &     0.310 \\
\bottomrule
\end{tabular}
\vspace{-4mm}
\end{table}

Table~\ref{table:cev_municipalities} describes the CEV for each municipality, and for all datasets, three components represent most of the variance, validating the use of a three-dimensional projection. Additionally, the three most important \textit{eigenprofiles} are plotted in Fig.~\ref{fig:muni_distributions}(a-c), showing a clear similarity pattern between them, meaning that elementary matrices (\ref{eq:elementary}) are similar. 

The most important observations are in the probability distributions for the spherical coordinates, shown in Fig.~{\ref{fig:muni_distributions}(d-f)} with their respective moments quantified in Table~\ref{table:statistics_sphere}. The distributions show that most data are concentrated in the sphere's shell. The radius distribution has negative skewness and mean closer to 1.0. Also, the data is concentrated in the polar angle ($\hat{\theta}$), which shows a single-mode distribution. The biggest difference between the distributions lies in the azimuthal angle with a higher standard deviation. This is expected because this angle covers the axes that keep most of the dataset variance, e.g., first and second \textit{eigenprofiles}, where most of the profile clusters are identified.

\section{Conclusion}


This paper presented a spherical model to represent load profiles using a PCA dimensionality reduction technique. The proposed modelling aids in the detection of outlier load profiles using a spherical coordinate system. The radius of the PCA projection is the most effective variable for detecting defective meters. The spherical profile visualisation uncovers a continuous underlying latent ordering within the load profiles. The ordering is carried out by applying a principal curve technique. A generative model was created by combining the principal curve and the von Fisher-Misses distribution from the directional statistics. Analyses of four different load profile datasets from different municipalities in the Netherlands show that the spherical structure with latent ordering exists, validating the spherical modelling for multiple datasets.

\vspace{-3mm}
\bibliographystyle{IEEEtran}
\bibliography{bibliography}

\end{document}